# Geometric Photonic Spin Hall Effect with Metapolarization

Xiaohui Ling[1,3],[#] Xinxing Zhou[2],[#] Xunong Yi[1], Hailu Luo[1,2],[*] and Shuangchun Wen[1,2][‡]

[1]*SZU-NUS Collaborative Innovation Center for Optoelectronic Science & Technology, and Key Laboratory of Optoelectronic Devices and Systems of Ministry of Education and Guangdong Province, Shenzhen University, Shenzhen 518060, China*

[2]*Laboratory for spin photonics, College of Physics and Microelectronic Science, Hunan University, Changsha 410082, China*

[3]*Department of Physics and Electronic Information Science, Hengyang Normal University, Hengyang 421002, China*

[*]**hailuluo@hnu.edu.cn**
[‡]**scwen@hnu.edu.cn**

[#]The two authors contributed equally to this work.

**Abstract**: We develop a geometric photonic spin Hall effect (PSHE) which manifests as spin-dependent shift in momentum space. It originates from an effective space-variant Pancharatnam-Berry (PB) phase created by artificially engineering the polarization distribution of the incident light. Unlikely the previously reported PSHE involving the light-matter interaction, the resulting spin-dependent splitting in the geometric PSHE is purely geometrically depend upon the polarization distribution of light which can be tailored by assembling its circular polarization basis with suitably magnitude and phase. This metapolarization idea enables us to manipulate the geometric PSHE by suitably tailoring the polarization geometry of light. Our scheme provides great flexibility in the design of various polarization geometry and polarization-dependent application, and can be extrapolated to other physical system, such as electron beam or atom beam, with the similar spin-orbit coupling underlying.

Angular momentum of light contains a spin part and an orbital part, playing an important role in light-matter interaction. The spin-orbit interaction of light is generally believed to be the origin of photonic spin Hall effect (PSHE), which describes the mutual influence of the spin (polarization) and the trajectory of light beam propagation [1–3]. It may result in two types of geometric phases, the Rytov-Vladimirskii-Berry phase associated with the evolution of the propagation direction of light and the Pancharatnam-Berry (PB) phase related to the manipulation with the polarization state of light [4–6]. The PSHE has been proved to hold great potential in spin photonics and advantageous metrology of material properties [3,7–9].

Recently, a new effect similar to the previously reported PSHE but not involve light-matter interaction, called geometric PSHE, is developed [11, 12]. This geometric effect has the same physical origin, i.e., the spin-orbit interaction, as the PSHE. The spin-orbit coupling for the PSHE can occur when a light beam reflects or refracts in a plane sharp interface [3, 10], where different angular spectrum components of the beam acquire different rotations of their polarization vectors, i.e., carrying different Rytov-Vladimirskii-Berry phase. Their interference results in the redistribution of the intensity and leads to the observation of spin-dependent shift of the beam centroids in the coordinate space. The spin-orbit coupling for the geometric PSHE can be observed when light beams propagating in free space with a titled observation plane [11-15]. Projecting the angular spectrum components onto the observation plane, their polarization vectors acquire different effective "rotations", and leads to the generation of effective, space-variant Rytov-Vladimirskii-Berry phase, and therefore the geometric PSHE. Until now, this interesting effect has been investigated for collimated paraxial beams [11], polarizing interfaces [12,13], orbital angular momentum carrying light beam [14], and tightly focused spin-segmented beams [15].

In this work, we develop a geometric PSHE by employing an effective space-variant PB phase which is associated with local polarization change of light. This kind of geometric phase produced by the light-matter interaction has been demonstrated in subwavelength gratings [16], liquid crystal $q$ plate [17], and plasmonic chains [18], etc. It usually indicates a spin-dependent splitting of light [18-20]. Here, we introduce an effective PB phase not depend upon the light-matter interaction but by artificially designing a light beam with transversely inhomogeneous polarization distribution rather than projecting the beam in a titled observation plane. In this case, the light beam acquires a space-variant PB phase while its propagation direction remained unchanged. The interference upon transmission may lead to the redistribution of the beam intensity and spin-dependent shift (angular shift) of the beam controids in the momentum space (*k* space). In fact, any polarization geometry can be obtained by engineering its circular polarization basis with suitable magnitudes and phase factors. This metapolarization idea enables us to manipulate the geometric PSHE by suitably engineering the polarization geometry of light.

Light beam with inhomogeneous polarization distribution can be viewed as a result of a beam with homogeneous polarization acquiring a locally varying polarization change, effectively, it carries a spin-dependent and space-variant PB phase [17,18]. The PB phase terms is $\exp(-i\sigma\Phi)$ with $\Phi$ representing a coordinate-dependent factor (typically, e.g., a vortex phase $\Phi=m\varphi$ with $m$ the topological charge and $\varphi$ the azimuthal angle in polar coordinate associated with the orbital angular momentum [20]), which indicates the coupling between the spin $\sigma$ and the orbit angular momentum. This produces a geometric PB phase gradient in coordinate space, and hence the momentum shift $\Delta k_{\sigma\pm}$. For the previously observed geometric PSHE, the Rytov-Vladimirskii-Berry phase term $\exp(-i\sigma k_y \tan\theta)$ is related to the coupling of the spin $\sigma$ and the transverse momentum $k_y$ [12]. Due to the different $k_y$ of the angular spectrum components, it forms a Rytov-Vladimirskii-Berry phase gradient in the momentum space, and therefore gives rise to a transverse shift $\Delta y$ in the coordinate space. This reveals the common physical origin, spin-orbit interaction, of the two kinds of geometric PSHE respectively related to the two types of geometric phases.

The polarization of light can be graphically represented by the so-called Poincaré sphere and algebraically described the following equation in terms of the azimuthal and polar angles $(\theta, \beta)$:

$$|\psi(\theta,\beta)\rangle = \cos\left(\frac{\theta}{2}\right)|R\rangle e^{i\beta/2} + \sin\left(\frac{\theta}{2}\right)|L\rangle e^{-i\beta/2}, \quad (1)$$

Here, |R> and |L> stand for the right- and left-circular polarizations, respectively, corresponding to the south ($\theta = 0$) and north poles ($\theta = \pi$) on the Poincaré sphere. It indicates that $\cos(\theta/2)$ and $\sin(\theta/2)$ are magnitude factors, while $e^{i\beta/2}$ and $e^{-i\beta/2}$ are phase factors. For $\theta = \pi/2$, Eq. (1) indicates a linear polarization on the equator of the Poincaré sphere with $\beta$ determining the longitude. Other values of $(\theta, \beta)$ represent elliptical polariations. Modifying the magnitude and phase factors, we can obtain any

desired polarization on the Poincaré sphere. If $β$ is not a constant, but a position-dependent function, the polarization becomes inhomogeneous.

We then consider cylindrical vector beam (CVB), which exhibits inhomogeneous linear polarization distribution and can be described by a Jones vector $(\cos β, \sin β)^T$ [21]. It can be decomposed into two orthogonally circular polarized bases with just opposite vortexes,

$$\begin{pmatrix} \cos β \\ \sin β \end{pmatrix} = \frac{1}{2}\left[ |R\rangle e^{-iβ} + |L\rangle e^{+iβ} \right], \qquad (2)$$

where $β=mφ+β_0$ with $β_0$ the initial polarization angle for $φ=0$. The CVBs can be graphically represented by the so-called Higher-order Poincaré sphere [see Fig. 1 (a) and 1(b)] [22, 23]. The two bases locate on the north and south poles, respectively. Their equal-weight superposition results in the linearly polarized CVBs on the equator, and elliptically polarized CVBs between the equator and the poles. This geometric description of the CVB is very similar to that of the fundamental Poincaré spheres for the homogeneous polarized light [24]. It is known that the linearly polarized CVBs does not produce the spin-dependent splitting in the free space, although the left- and right-handed components of the linearly polarized CVB exhibit an effective, spin-dependent PB phase. This is due to that the polarization of this kind of beam possesses rotational symmetry, and forms a uniform polarization rotation rate in the azimuthal direction, so the left- and right-handed components always superpose exactly at the same position, which still manifests as a linearly polarized CVB.

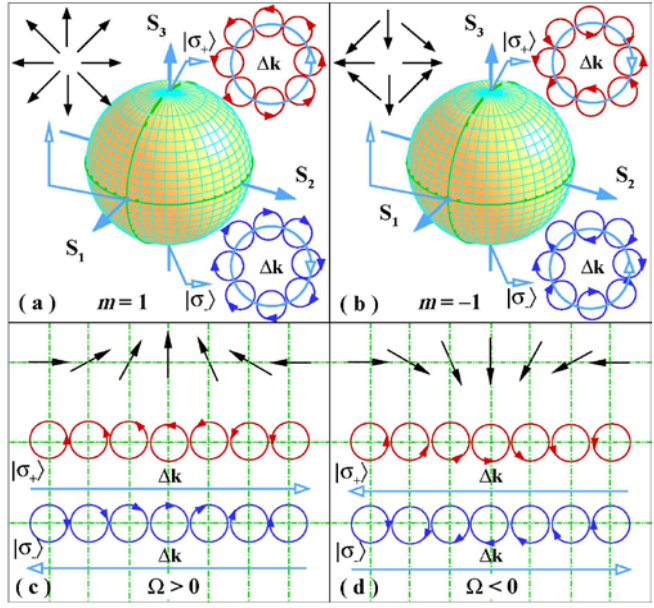

FIG. 1: (Color online) Schematic pictures of the vector beam. The CVB can be represented by the Higher-order Poincaré sphere for (a) $m=1$ and (b) $m=-1$. Transforming the CVB from the polar coordinate to the Cartesian coordinate, the linearly polarized CVB in the equator unfolds its polarization with polarization rotation rate in the $x$ direction for (c) $\Omega > 0$ and (d) $\Omega < 0$.

Breaking the rotational symmetry of the CVBs, it is possible to observe the spin-dependent splitting and momentum shift. A birefringent wave plate can be employed to break the rotational symmetry [25], or unfolding the CVB from polar coordinate to the Cartesian coordinate [26]. This polarization-unfolded vector beam possesses a polarization rotation rate in the $x$ or $y$ direction [Fig. 1 (c) and (d)], and carries a spin-dependent phase which can be viewed as an effective PB phase. In this case, the local polarization direction is written as a coordinate-dependent function $\beta=\Omega x+\beta_0$ with $\Omega=2\pi/d$ the uniform polarization rotation rate. Here, $d$ is the period of the polarization variation. So the spin-dependent momentum shift can be obtained as the gradient of the PB phase ($\Phi_{PB}=\sigma_{\pm}\beta$), that is, $\Delta k_{\sigma\pm}=\nabla\Phi_{PB}=\sigma_{\pm}\Omega e_x$, where $\sigma_{\pm}=\pm 1$ represents the left-and right-handed circular polarization and $e_x$ the unit vector in the $x$ direction. The induced real-space shift upon transmission can be calculated as $\Delta x_{\sigma\pm}=(z/k_0)\Delta k_{\sigma\pm}$ with $k_0=2\pi/\lambda$ the free-space wave number and $z$ the transmission distance (see Fig. 2 for the mapping relation between the momentum shift and the real-space shift).

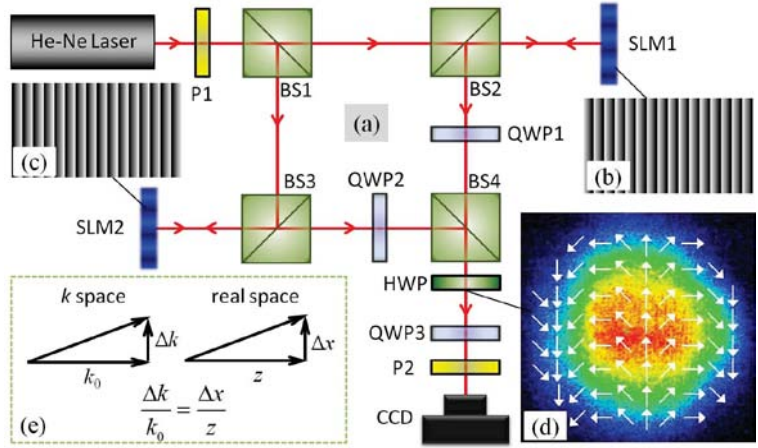

FIG. 2: (Color online) (a) Experimental setup for generating the polarization-unfolded beam shown in Fig. 1(d). The collimated He-Ne laser (632.8 nm, 21mW) produces a linearly polarized Gaussian beam and the polarizer (P1) ensures the polarization direction be horizontal (*x* direction). The four beam splitters (BS) form a Mach-Zender interferometer. The two spatial light modulators (SLM, Holoeye Pluto-Vis, Germany) apply just opposite phase gradients in the *x* direction [see the insets (b) and (c)] to the two sub-beams, respectively. And two quarter-wave plates (QWP) inclined to 45 and -45 to the *x* direction, respectively, produce two circular polarizations with opposite handedness before they superpose in the BS4. After the BS4, the beam possesses a polarization rotation rate in the *x* direction; see the schematic picture in the inset (d). A half-wave plate (HWP) can be used to reverse the polarization rotation rate. A circular polarizer (QWP3 and a polarizer P2) and a CCD camera are used to measure the Stokes parameter $S_3$. (e) The mapping relation between the momentum shift $\Delta k$ and the induced real-space shift $\Delta x$.

To measure the geometric PSHE, we set up a Mach-Zender interferometer constructed by four beam splitters (BS) and two quarter-wave plates (QWP) to generate the polarization-unfolded vector beam (Fig. 2). A collimated He-Ne laser with a fundamental Gaussian profile is split into two sub-beams, then impinging onto two reflective phase-only spatial light modulators (SLM), respectively. The SLMs can apply just opposite phase modulation to the two sub-beams by engineering the phase patterns [see the insets (b) and (c) in the figure] displayed on the SLMs. And then the two sub-beams are justified as circular polarization with the desired handedness by means of properly oriented QWPs. Before superposing at the BS4, they respectively indicate the two terms in the right hand of the Eq. (1), i.e., two orthogonal circular polarizations with just opposite dynamic phase modulation. After superposition, the resultant beam is the desired polarization-unfolded vector beam exhibiting a spatial polarization rotational rate in the *x* direction. It is note that this beam is not stable upon transmission because the effective PB phase results in a spin-dependent momentum shift, i.e., the two circular polarizations will separate from each other upon transmission. Our experimental setup shows an effective method to mimic the geometric phase by tailoring the dynamic phase in the SLMs.

Additionally, a half-wave plate (HWP) followed with the BS4 can be employed to reverse the polarization rotation rate, because a linearly polarized beam with Jones vector $(\cos\beta, \sin\beta)^T$ can be transformed to a beam with Jones vector $[\cos(2\psi-\beta), \sin(2\psi-\beta)]^T$ with $\psi$ the optical axis direction of the HWP. For $\psi=0$, the polarization rotation rate just reverses. The spin-dependent splitting can be demonstrated by the Stokes parameter $S_3$ because the $S_3$ can describe the circular polarization degree. We can measure this parameter by a typical setup, a quarter-wave plate (QWP3), a polarizer (P2), and a CCD camera.

We now measure the real-space shift $\Delta x$, which is the coherent superposition contributed by the local momentum shift. As the polarization rotation rate is actually formed by the dynamic phase gradient created by the SLM, we can construct different phase gradient via defining a pixel (8 μm×8 μm, the data is from the manual of the SLM) of the SLM for a different phase level. Each phase level is $2\pi/N$ radian with $N$ the number of the total phase level in one period's variation. So the period $d$ of the polarization variation is $8\times N$ μm. We first consider the influence of the transmission distance $z$ on the geometric PSHE. As the shift in momentum space is essential an angular shift, the induced real-space shift increases with $z$ linearly [Fig. 3(a)].

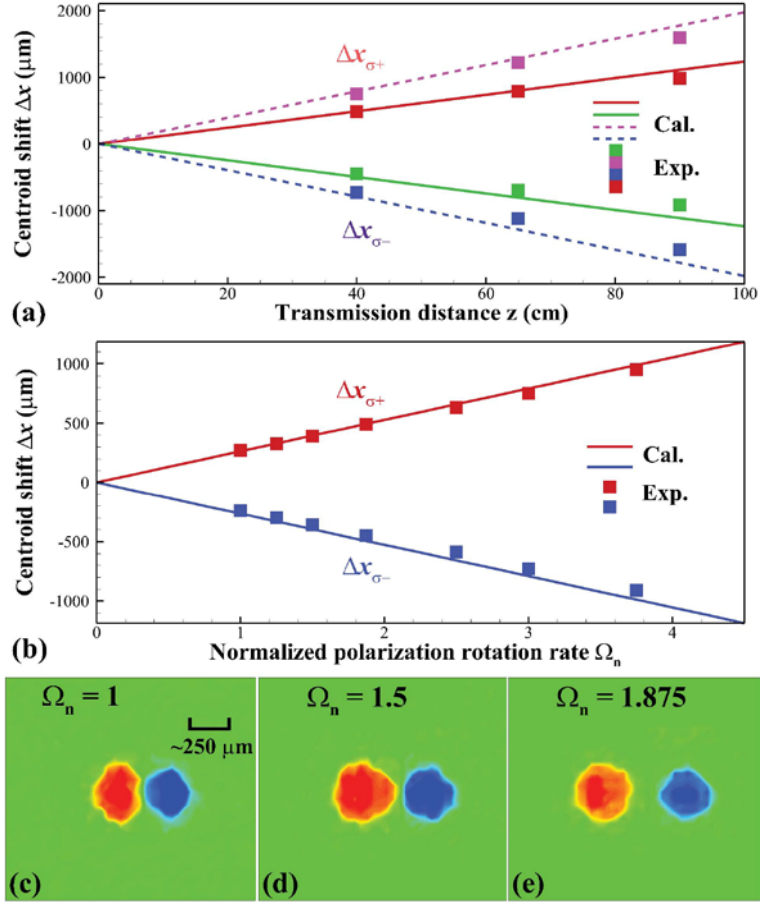

FIG. 3: (Color online) (a) The calculated (Cal.) and experimentally (Exp.) measured real-space shifts for different transmission distance (40 cm, 65 cm, and 90 cm) for two given polarization rotation rates. For the sake of simplicity, the polarization rotation rate is normalized as $\Omega_n=\Omega\times 960/(2\pi)$. The solid lines indicate the results for $\Omega_n=1.5$ and the dashed lines for $\Omega_n=1.875$. (b) The calculated and measured real-space shifts for different polarization rotation rates in a given transmission distance (40 cm away from the BS4). (c)-(e) Examples of the circular polarization degree (Stokes parameter $S_3$) of the resulting beam for different $\Omega_n$.

The phase level is inversely proportional to the value of $N$. When $N$ increases, the phase level and the polarization rotation rate become smaller. In fact the real-space shift also increases linearly with the polarization rotation rate [Fig. 3(b)]. We carry out the experiment for several polarizations rotation rates. These results agree well with the theoretical ones. Several examples of the $S_3$ of the resulting beam for different polarization rotation rate are shown in Fig. 3(c).

Suitably engineering the polarization distribution of the incident light beam, we can modulate the spin-dependent splitting and the geometric PSHE. We now design a polarization-segmented light beam with different polarization rotation rates for the upper (left) and bottom (right) portion of the beam [see the schematic picture in Fig. 4(a) and (d)]. This can be achieved by suitably programming the phase patterns displayed on the SLMs. Due to their different polarization rotation rates, the two parts produce different shifts, so it manifests as four splitting lobes. When the two polarization rotation rates have opposite signs, the $\sigma_+$ or $\sigma_-$ component shift to the inverse directions [see Fig. 4(b) and 4(c)]. In fact if the polarization rotation rate exists in the $y$ direction, the corresponding shifts are also produced in this direction [see Fig. 4(e) and 4(f)].

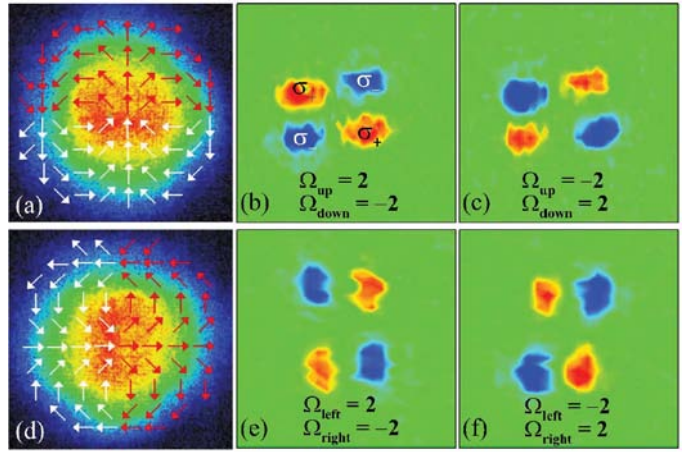

FIG. 4: (Color online) (a) Schematic picture of the polarization-segmented vector beam. The upper part exhibits an opposite polarization rotation rate with the bottom part. The arrows indicate the local polarization directions of the beam. (b) and (c) shows the spin-dependent splitting ($S_3$ parameter) for two polarization-segmented beams with just the opposite polarization rotation rates. (d) Schematic picture of a polarization-segmented vector beam with polarization rotation rate in the $y$ direction. (e) and (f) are the examples of the spin-dependent splitting of (d).

In conclusion, we have demonstrated a new kind of geometric PSHE which manifests itself as spin-dependent splitting in momentum space based on the effective PB phase by employing the artificial metapolarization. This effect has the similar physical origin, i.e., effective spin-orbit coupling, as the previously reported geometric PSHE. Suitably engineering the polarization distribution of light beam, we can mimic the space-variant PB phase with an interferometer and therefore modulating the geometric PSHE. We use a polarization-unfolded vector beam which can be viewed as result of the transformation of a CVB from polar coordinate to the Cartesian coordinate to verify the idea. As polarization is a fundamental property of light, this generalized geometric PSHE achieved by polarization-engineered metapolarization may play an important role in future spin-based photonics applications.

This research was supported by the National Natural Science Foundation of China (Grants No. 11274106 and No. 11347120), the Scientific Research Fund of Hunan Provincial Education Department of China (Grant No. 13B003), and the Doctorial Start-up Fund of Hengyang Normal University (Grant No. 13B42).